%% file: arxiv.tex
\documentclass{article}
\pdfoutput=1
\usepackage{lamperskipackages}
\usepackage{lamperskicommands}
\usepackage{lamperskitheorem}
\author{Andrew Lamperski and John C. Doyle}
\title{Output Feedback $\mathcal{H}_2$ Model Matching for Decentralized Systems with Delays} 
\begin{document}
\maketitle
\input{abstract}
\input{intro}
\input{problem}
\input{centralized}
\input{decentralized}
\input{examples}

\input{conclusion}
\bibliography{bibliography}
\end{document}

%% file: abstract.tex
\begin{abstract}
This paper gives a new solution to the output feedback $\Htwo$
model matching problem for a large class of delayed 
information sharing patterns. Existing methods for such problems
typically reduce the decentralized problem to a centralized problem of
higher state dimension. In contrast, the controller given in this
paper is constructed from the solutions to the centralized control and
estimation Riccati equations for the original system.  
The problem is solved by decomposing the 
controller into two components. One is centralized, but delayed, while
the other is decentralized with finite impulse response (FIR). It is
then shown that the optimal controller can be constructed through a
combination of centralized spectral factorization and quadratic programming.
\end{abstract}

%% file: intro.tex
\section{Introduction}

Decentralized control problems arise when inputs to a dynamic
system are chosen by multiple controllers with access to different
information.  In decentralized control with delays,
local measurements are passed to the various controllers
over a communication network with delays. As a result of the
delays, some controllers will have access to measurements before
others. This paper provides a new solution to the $\Htwo$ model
matching problem, subject to communication delays, based on spectral
factorization. 


\subsection{Related Work}

A large number of dynamic programming methods have been developed for
decentralized optimal control problems.   For the special case
known as the one-step delay information sharing pattern, the output
feedback $\Htwo$
problem was solved in the 1970s by dynamic programming 
\cite{sandellsolution1974,kurtaranlinearquadraticgaussian1974,yoshikawadynamic1975}. 
For more complex delay patterns, dynamic programming has extensions to decentralized
state feedback
\cite{aicardidecentralized1987,lamperskistructure2011,lamperskidynamic2012},
but output feedback 
is difficult
because the separation principle
fails
\cite{varaiyadelayed1978,yoshikawaseparation1978,kurtarancorrections1979}. 
Recently, methods based on POMDPs have been developed for output
feedback control of
nonlinear systems with general delay patterns
\cite{nayyaroptimal2011,nayyardecentralized2012}. 

In the past few years, spectral factorization has been employed to
derive explicit solutions to the $\Htwo$ problem with
sparsity constraints, but not delays. First, decentralized state
feedback was addressed 
\cite{swigartexplicit2010a,shahh2optimal2010}, followed by restricted types of
output feedback \cite{swigartoptimal2011, kimunifying2011} , and most
recently full output feedback \cite{lessardoptimal2012}. In 
these works, it was shown how to efficiently construct decentralized solutions
using standard Riccati equations. This paper applies
spectral factorization to delayed information sharing patterns. As in
the sparsity constrained case, the resulting controllers are
efficiently computable from solutions to centralized Riccati equations.

\subsection{Existing Solutions}

The output feedback $\Htwo$ problem with communication delays, as
studied in this paper, has been
previously solved using approaches
based on vectorization \cite{rotkowitzcharacterization2006}, linear
matrix inequalities (LMIs) 
\cite{rantzerseparation2006,gattamigeneralized2006}. The problem can
also be solved as a special case of the work in
\cite{nayyardecentralized2012}.
All of these solutions reduce the decentralized control problem to a
centralized problem of higher state dimension. 


\subsection{Contributions}

The main contribution of this paper is a novel efficient solution to a 
general class of decentralized $\Htwo$ output feedback model matching problems
with communication 
delays.  Unlike the existing approaches mentioned above,  the method
of this paper works directly with the original state matrices. In fact, the solution is constructed from the classical
control and estimation Riccati equations for the original system.

A key assumption made in this paper is that each local measurement
eventually reaches each controller. This assumption allows the
controller to be decomposed into a centralized, but delayed, component
and a decentralized finite 
impulse response (FIR) component. Similar decompositions have been exploited in
\cite{yoshikawadynamic1975,lamperskistructure2011,lamperskidynamic2012,nayyaroptimal2011,nayyardecentralized2012,rantzerlinear2006}. 
Given the decomposition, the optimal
centralized component can be computed as a function of the FIR
component by a relatively straightforward extension of centralized
spectral factorization. It is
then shown that the optimal FIR component can be found by quadratic
programming.

In the case of a quadratically invariant delay pattern \cite{rotkowitzcharacterization2006},
optimal decentralized feedback controllers can be computed via the
Youla parametrization. When the delay pattern is not  
quadratically invariant, the model matching procedure of this paper 
is still optimal, but the feedback controller recovered by a linear
fractional transformation is not guaranteed to satisfy the delay constraint.

\subsection{Overview}

The paper is structured as follows. Section \ref{sec:problem}
defines the general problem studied in this paper. Section
\ref{sec:centralized} reviews spectral factorization for centralized
$\Htwo$ model matching in both undelayed and delayed cases. Extending
the delayed centralized model matching technique, the decentralized
problem is solved in Section \ref{sec:decentralized}. Numerical
results are given in Section \ref{sec:examples} and finally conclusions are
given in \ref{sec:conclusion}.

%% file: problem.tex
\section{Problem}
\label{sec:problem}

This section introduces the basic notation and the model matching
problem of interest. Subsection \ref{sec:communication} describes how
common delayed information sharing patterns can be cast in the
framework of this paper.  

\subsection{Preliminaries on $\Htwo$}
Let $\D = \{z\in \C:|z|<1\}$ be the unit disc of complex numbers. A function $G:(\C\cup\{\infty\})\setminus \D\to \C^{p\times q}$ is in $\Htwo$ if it can be expanded as 
\[
G(z) = \sum_{i=0}^{\infty}\frac{1}{z^i}G_i,
\]
where $G_i\in \C^{p\times q}$ and $\sum_{i=0}^{\infty}\Tr(G_iG_i^T)<\infty$. Define the conjugate of $G$ by 
\[
G(z)^{\sim} = \sum_{i=0}^{\infty}z^iG_i^*.
\]
For a real rational transfer matrix, $G = \left[\begin{array}{c|c}
A & B \\
\hline 
C & D
\end{array}
\right]
$, the conjugate is given by 
\[
\left(C(zI-A)^{-1}B+D\right)^{\sim} = 
B^T\left(
\frac{1}{z}I-A^T
\right)^{-1}C^T+D^T.
\]

The space $\Htwo$ is a Hilbert space with inner product defined by 
\begin{eqnarray*}
\langle G,H\rangle &=&
 \frac{1}{2\pi}\int_{-\pi}^{\pi}\Tr\left(
G\left(e^{j\theta}\right)H\left(e^{j\theta}\right)^{\sim} 
\right)
 d\theta \\
&=& \sum_{i=0}^{\infty} \Tr\left(
G_iH_i^*
\right),
\end{eqnarray*}
where the second equality follows from Parseval's identity. 

If $\mathcal{M}$ is a subspace of $\Htwo$, denote the orthogonal
projection onto $\mathcal{M}$ by $\P_{\mathcal{M}}$.


\subsection{Formulation}

This subsection introduces the generic problem of interest. Let $P$ be a stable discrete-time plant given by 
\[
P = 
\left[
\begin{array}{c|cc}
A & B_1 & B_2 \\
\hline
C_1 & 0 & D_{12} \\
C_2 & D_{21} & 0
\end{array}
\right] = 
\begin{bmatrix}
P_{11} & P_{12} \\
P_{21} & P_{22}
\end{bmatrix},
\]
with inputs of dimension $p_1$, $p_2$ and outputs of dimension $q_1$, $q_2$.
Attention will be restricted to stable plants for simplicity. Unstable
plants can be handled by first applying a stabilizing feedback and
optimizing the resulting system. 

For the existence of solutions of the appropriate Riccati equations,
assume that 
\begin{itemize}
\item $D_{12}^TD_{12}$ is positive definite,
\item $(A,B_1)$ is stabilizable,
\item $D_{21}D_{21}^T$ is positive definite,
\item $(C_1,A)$ is detectable.
\end{itemize}
(Note that stabilizability and detectability follow immediately from
the stability assumption.)

For $N\ge 1$, define the space of strictly proper finite impulse
response (FIR) transfer matrices by $\mathcal{X} =
\bigoplus_{i=1}^N\frac{1}{z^i}\C^{p_2\times q_2}$. 
Note that 
$\frac{1}{z}\Htwo$ 
can be decomposed into orthogonal subspaces as 
\begin{eqnarray*}
\frac{1}{z}\Htwo &=& \mathcal{X} \oplus \frac{1}{z^{N+1}} \Htwo, 
\end{eqnarray*}

Let $\RRp$ be the space of proper real rational transfer matrices. 
Let $\mathcal{S}\subset\frac{1}{z}\mathcal{R}_{\mathrm{p}}$
be a subspace of 
the form 
\begin{eqnarray}\label{eq:subspace}
\mathcal{S} &=&  
\mathcal{Y}\oplus\frac{1}{z^{N+1}}\mathcal{R}_{\mathrm{p}},
\end{eqnarray}
where $\mathcal{Y}\subset \bigoplus_{i=1}^N \frac{1}{z^i}\R^{p_2\times
  q_2} \subset \mathcal{X}$. 

\begin{figure}
\centering
\includegraphics[width=.4\columnwidth]{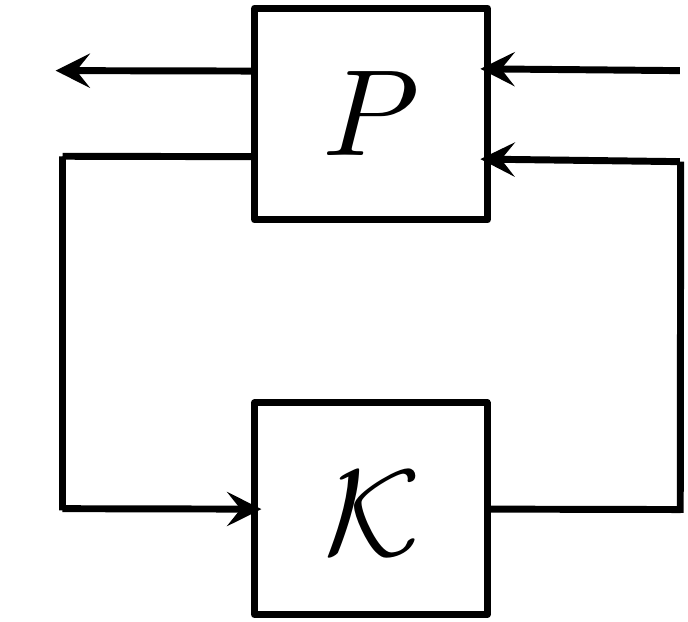}
\caption{\label{fig:loop}
The basic feedback loop
}
\end{figure}

\gap

The decentralized $\Htwo$ model matching problem considered in this
paper is given by 
\begin{equation}\label{eq:modelMatching}
\begin{array}{c}
\min_{Q} \|P_{11}+P_{12}QP_{21}\|_{\Htwo} \\
\textrm{s.t. } Q\in \mathcal{S} \cap \frac{1}{z}\Htwo.
\end{array}
\end{equation}

A feedback controller for the plant can be defined from $Q$ by $\mathcal{K} =
Q(I+P_{22}Q)^{-1}$. If
the space $\mathcal{S}$ is quadratically invariant,\footnote{The space
  $\mathcal{S}$ is quadratically invariant if
  $\mathcal{K}P_{22}\mathcal{K}\in \mathcal{S}$ for all
  $\mathcal{K}\in \mathcal{S}$.} then $Q\in \mathcal{S}$ if and only if
$\mathcal{K}\in\mathcal{S}$,
\cite{rotkowitzcharacterization2006}. Furthermore, since $Q$ 
is strictly proper and stable, and $P$ is stable, $\mathcal{K}$ must be strictly proper
and the closed loop system
$P_{11}+P_{12}\mathcal{K}(I-P_{22}\mathcal{K})^{-1}P_{21}$ must be 
stable. Furthermore, if $Q$ solves the model matching problem, then
$\mathcal{K}$ must solve the decentralized optimal control problem:
\begin{equation}\label{eq:standardProb}
\begin{array}{c}
\min_{\mathcal{K}}
 \|P_{11}+P_{12}\mathcal{K}(I-P_{22}\mathcal{K})^{-1}P_{21}\|_{\Htwo} \\
\textrm{s.t. } \mathcal{K}\in \mathcal{S}.
\end{array}
\end{equation}

Note that even if $\mathcal{S}$ is not quadratically invariant,
the model matching problem is still convex, and can
be solved by the methods in this paper. In this case, however, it
could happen that $Q(I+P_{22}Q)^{-1}\notin \mathcal{S}$, and
thus the solution to Problem 
(\ref{eq:modelMatching}) need not lead to a solution of Problem
(\ref{eq:standardProb}).

For technical simplicity, controllers in this paper are assumed to be
strictly proper (that is, in
$\frac{1}{z}\mathcal{R}_{\mathrm{p}}$). The results in this paper can
be extended to non-strictly proper controllers but more complicated
formulas would result. 

\subsection{Communication Delay Patterns}

\label{sec:communication}

Equation (\ref{eq:subspace}) can be used to model many delayed
information sharing patterns. For instance, an infinite-horizon, strictly
proper version of the $1$-step delayed information sharing pattern studied in
\cite{sandellsolution1974,kurtaranlinearquadraticgaussian1974,yoshikawadynamic1975} 
is captured by the case that $N=1$ and $\mathcal{Y}$ corresponds to
block diagonal FIR matrices
\[
\mathcal{Y} = \frac{1}{z} 
\begin{bmatrix}
\R^{p_{21}\times q_{21}} & 0 \\
0 & \R^{p_{22}\times q_{22}}
\end{bmatrix}.
\]
Similarly, for $N>1$, the $N$-step delay information sharing pattern
studied  
in
\cite{varaiyadelayed1978,yoshikawaseparation1978,kurtarancorrections1979,nayyaroptimal2011}
can be characterized by $\mathcal{Y}$ of the form 
\[
\mathcal{Y} = \bigoplus_{i=1}^{N} \frac{1}{z^i} 
\begin{bmatrix}
\R^{p_{21}\times q_{21}} & 0 \\
0 & \R^{p_{22} \times q_{22}}
\end{bmatrix}.
\] 

More general team problems with communication delays,
such those studied in
\cite{lamperskistructure2011,lamperskidynamic2012,nayyardecentralized2012,rantzerseparation2006,gattamigeneralized2006,rantzerlinear2006},
can also be captured by Equation (\ref{eq:subspace}).
For instance, a
strictly proper version of the three-player chain problem discussed in \cite{lamperskistructure2011} is
described by $N=2$ and
\begin{equation} \label{eq:chain}
\mathcal{Y}_{\mathrm{Ch}} = \frac{1}{z}
\begin{bmatrix}
* & 0 & 0 \\
0 & * & 0 \\
0 & 0 & *
\end{bmatrix}\oplus
\frac{1}{z^2}
\begin{bmatrix}
* & * & 0 \\
* & * & * \\
0 & * & *
\end{bmatrix},
\end{equation}
where, for compactness, $*$ is used to denote a space of appropriately
sized real matrices. 

%% file: centralized.tex
\section{Centralized Spectral Factorization}
\label{sec:centralized}

This section gives spectral factorization solutions to centralized
model matching problem in both delayed and undelayed cases. While the
solutions are classical, 
they will be presented in detail, as the decentralized
model matching problem relies heavily on the terms and ideas in the centralized
solutions.  

\subsection{Undelayed Case}

The undelayed case corresponds to 
\begin{equation}\label{eq:modelMatchingNoDelay}
\begin{array}{c}
\min_Q \|P_{11}+P_{12}QP_{21}\|_{\Htwo} \\
\textrm{s.t. } Q\in \frac{1}{z}\Htwo.
\end{array}
\end{equation}

A necessary condition for optimality is given by 
\begin{equation}\label{eq:optimalityNoDelay}
P_{12}^{\sim}P_{11}P_{21}^{\sim}+P_{12}^{\sim}P_{12}QP_{21}P_{21}^{\sim}
\in \left(\frac{1}{z}\Htwo\right)^{\perp}.
\end{equation}
A simple argument shows that $\left(\frac{1}{z}\Htwo\right)^{\perp} =
\frac{1}{z}\Htwo^{\perp}$, a fact that will be used several times. 

To derive the optimality condition, 
let $\delta\in\frac{1}{z}\Htwo$ be a small perturbation. The perturbed norm can
be expanded as
\begin{eqnarray*}
\lefteqn{
\|P_{11}+P_{12}(Q+\delta)P_{21}\|_{\Htwo}^2}
\\
 &=& 
\langle P_{11}+P_{12}(Q+\delta)P_{21},P_{11}+P_{12}(Q+\delta)P_{21}
\rangle
\\
&=&
\|P_{11}+P_{12}QP_{21}\|_{\Htwo}^2+
\\ &&
2\mathrm{Re}\langle
P_{12}^{\sim}(P_{11}+P_{12}QP_{21})P_{21}^{\sim},\delta
\rangle
+O(\|\delta\|_{\Htwo}^2).
\end{eqnarray*}
In particular, if $Q$ is optimal, then the second term must
vanish for any $\delta$, and it follows that Equation
(\ref{eq:optimalityNoDelay}) must hold. 

The following classical lemmas show how to factorize
$P_{12}^{\sim}P_{12}$ and $P_{21}P_{21}^{\sim}$ as products of
causally invertible transfer matrices. 

\gap

\begin{lemma}
{\it
Let $X$ be the stabilizing solution of the Riccati equation
\begin{eqnarray*}\label{eq:controlRiccati}
X &=&
C_1^TC_1+A^TXA-
\\ && \nonumber
(A^TXB_2+C_1^TD_{12})\Omega^{-1}
(B_2^TXA+D_{12}^TC_1),
\end{eqnarray*}
where $\Omega = D_{12}^TD_{12}+B_2^TXB_2$. Define the (linear quadratic
regulator) gain by  
\[
K = -\Omega^{-1}(B_2^TXA+D_{12}^TC_1).
\]
The transfer matrix $P_{12}^{\sim}P_{12}$ has a left spectral
factorization $P_{12}^{\sim}P_{12}=W_L^{-\sim}W_L^{-1}$, where $W_L$
is given by 
\begin{eqnarray*}
W_L&=&\left[
\begin{array}{c|c}
A+B_2K & B_2 \\
\hline
K & I
\end{array}
\right]
\Omega^{-1/2}
,
\\
W_L^{-1}&=&\Omega^{1/2}
\left[
\begin{array}{c|c}
A& -B_2 \\
\hline
K & I
\end{array}
\right].
\end{eqnarray*}
}
\end{lemma}

\gap

\begin{lemma}
{\it
Let $Y$ be the stabilizing solution of the Riccati equation
\begin{eqnarray*}\label{eq:estimationRiccati}
Y&=&B_1B_1^T+AYA^T-
\\&& \nonumber
(AYC_2^T+B_1D_{21}^T)\Psi^{-1}
(C_2YA^T+D_{21}B_1^T),
\end{eqnarray*}
where $\Psi = D_{21}D_{21}^T+C_2YC_2^T$. Define the (Kalman filter) gain by
\[
L = -(AYC_2^T+B_1D_{21}^T)\Psi^{-1}.
\]
The transfer matrix $P_{21}P_{21}^{\sim}$ has a right spectral
factorization $P_{21}P_{21}^{\sim}=W_R^{-1}W_R^{-\sim}$ where $W_R$ is
given by 
\begin{eqnarray*}
W_R&=&\Psi^{-1/2}
\left[
\begin{array}{c|c}
A+LC_2& L \\
\hline
C_2 & I
\end{array}
\right]
,
\\ 
W_R^{-1}&=&
\left[
\begin{array}{c|c}
A & L \\
\hline
-C_2 & I
\end{array}
\right]
\Psi^{1/2}
\end{eqnarray*}
}
\end{lemma}

\gap

The following standard theorem gives the spectral factorization
solution to the model matching problem. The presentation is slightly
non-standard, in that the optimal matrix $Q_0$ is defined in terms
of an auxiliary matrix $T$, which is used in the delayed and decentralized
solutions.

\gap

\begin{theorem}
{\it
Define $T$ by 
\begin{equation}\label{eq:T}
T=\Omega^{1/2}\left[
\begin{array}{c|c}
A & L \\
\hline
K & 0
\end{array}
\right]
\Psi^{1/2}.
\end{equation}
The optimal solution to the model matching problem of Equation
(\ref{eq:modelMatchingNoDelay}) is given by
\[
Q_0 = -W_LTW_R.
\]
}
\end{theorem}

\gap

\begin{proof}
Assume that Equation (\ref{eq:optimalityNoDelay}) holds. 
Plugging in the spectral factorizations shows that 
\[
P_{12}^{\sim}P_{11}P_{21}^{\sim}+W_L^{-\sim}W_L^{-1}QW_R^{-1}W_R^{-\sim}
\in \frac{1}{z}\Htwo^{\perp}. 
\]
Anticausality of $W_L^{\sim}$ and $W_R^{\sim}$ implies that 
\begin{equation}
\label{eq:cenSimpPerp}
W_L^{\sim}P_{12}^{\sim}P_{11}P_{21}^{\sim}W_R^{\sim}+W_L^{-1}QW_R^{-1}
\in \frac{1}{z}\Htwo^{\perp}. 
\end{equation}
Note that $W_L^{-1}QW_R^{-1}\in  \frac{1}{z}\Htwo$. It follows that Equation
(\ref{eq:cenSimpPerp}) can be set to zero by applying the projection operator:
\begin{eqnarray}
\nonumber 
\lefteqn{
\P_{\frac{1}{z}\Htwo}\left(
W_L^{\sim}P_{12}^{\sim}P_{11}P_{21}^{\sim}W_R^{\sim}+W_L^{-1}QW_R^{-1}
\right)
}
\\ &=& \nonumber 
\P_{\frac{1}{z}\Htwo}\left(W_L^{\sim}P_{12}^{\sim}P_{11}P_{21}^{\sim}W_R^{\sim}\right)+
W_L^{-1}QW_R^{-1} 
\\ &=& 0. \label{eq:cenQT}
\end{eqnarray}
Let
$T=\P_{\frac{1}{z}\Htwo}\left(W_L^{\sim}P_{12}^{\sim}P_{11}P_{21}^{\sim}W_R^{\sim}\right)$,
Equation (\ref{eq:cenQT}) shows that $Q = -W_LTW_R$. Furthermore,
standard state space manipulations show that $T$ has the
form in Equation (\ref{eq:T}), and the proof is complete. 
\end{proof}

\subsection{Delayed Case}

The delayed case corresponds to the following model matching problem:
\begin{equation}\label{eq:modelMatchingDelayed}
\begin{array}{c}
\min_{Q}\|P_{11}+P_{12}QP_{21}\|_{\Htwo} \\
\textrm{s.t. }Q\in\frac{1}{z^{N+1}}\Htwo.
\end{array}
\end{equation}

An argument analogous to the derivation of Equation
(\ref{eq:optimalityNoDelay}) shows that a necessary condition for
optimality in the delayed case is 
\begin{equation}\label{eq:optimalityDelayed}
P_{12}^{\sim}P_{11}P_{21}^{\sim}+P_{12}^{\sim}P_{12}QP_{21}P_{21}^{\sim}\in 
\left(\frac{1}{z^{N+1}}\Htwo\right)^{\perp}.
\end{equation}
As in the undelayed case, a simple argument shows that 
$\left(\frac{1}{z^{N+1}}\Htwo\right)^{\perp}=\frac{1}{z^{N+1}}\Htwo^{\perp}$. 

\gap

\begin{theorem}\label{thm:centralizedDelayed}
{\it
The optimal solution to the delayed model matching problem is given by 
\begin{equation}\label{eq:solutionDelayed}
Q_N = -W_L\P_{\frac{1}{z^{N+1}}\Htwo}(T)W_R.
\end{equation}
}
\end{theorem}

\gap

\begin{proof}
Assume that $Q\in \frac{1}{z^N}\Htwo$ satisfies Equation
(\ref{eq:optimalityDelayed}) and thus 
\begin{eqnarray}
\hspace{-15pt}
\nonumber
P_{12}^{\sim}P_{11}P_{21}^{\sim}+W_L^{-\sim}W_L^{-1}QW_R^{-1}W_R^{-\sim}&\in&
\frac{1}{z^{N+1}}\Htwo^{\perp}, \\
\label{eq:delCenSimp}
W_L^{\sim}P_{12}^{\sim}P_{11}P_{21}^{\sim}W_R^{\sim}+W_L^{-1}QW_R^{-1} 
&\in& \frac{1}{z^{N+1}}\Htwo^{\perp},
\end{eqnarray}
where the second line follows from anticausality of $W_L^{\sim}$ and
$W_R^{\sim}$. 
As in the proof of the case with no delays, $W_L^{-1}QW_R^{-1}\in
\frac{1}{z^{N+1}}\Htwo$ and the left side of Equation
(\ref{eq:delCenSimp}) can be set to zero by projection:
\begin{eqnarray}
\nonumber
\lefteqn{
\P_{\frac{1}{z^{N+1}}\Htwo}\left(
W_L^{\sim}P_{12}^{\sim}P_{11}P_{21}^{\sim}W_R^{\sim}+
W_L^{-1}QW_R^{-1} 
\right) }
\\ &=& 
\nonumber
\P_{\frac{1}{z^{N+1}}\Htwo}\left(
W_L^{\sim}P_{12}^{\sim}P_{11}P_{21}^{\sim}W_R^{\sim}\right)
+
W_L^{-1}QW_R^{-1} 
\\ &=& 0. \nonumber\label{eq:delQT}
\end{eqnarray}
Furthermore, since
$\frac{1}{z^{N+1}}\Htwo\subset \frac{1}{z}\Htwo$, it follows that
$\P_{\frac{1}{z^{N+1}}\Htwo} = \P_{\frac{1}{z^{N+1}}\Htwo}\P_{\frac{1}{z}\Htwo}$. Thus,
the projection can be computed in terms of $T$ as
\begin{eqnarray*}
W_L^{-1}QW_R^{-1} &=& -\P_{\frac{1}{z^{N+1}}\Htwo}\left(
\P_{\frac{1}{z}\Htwo}\left(
W_L^{\sim}P_{12}^{\sim}P_{11}P_{21}^{\sim}W_R^{\sim}
\right)
\right) \\
&=&
-\P_{\frac{1}{z^{N+1}}\Htwo}(T).
\end{eqnarray*}
Multiplying on the left and right by $W_L$ and $W_R$, respectively,
completes the proof.
\end{proof}

%% file: decentralized.tex
\section{Decentralized Model Matching}
\label{sec:decentralized}

This section presents the main results of the paper. 
Recall that in centralized model matching, from Equation
(\ref{eq:modelMatching}), that $Q$ is constrained to be in the space
$\frac{1}{z^{N+1}}\Htwo\oplus \mathcal{Y}$. It follows that without loss of generality, $Q$ can be
decomposed as 
\[
Q = U+V
\]
with $U\in \frac{1}{z^{N+1}}\Htwo$ and $V\in \mathcal{Y}$. 

\gap

\begin{theorem}\label{thm:main}
{\it The optimal solution to the decentralized  
model matching problem (Equation (\ref{eq:modelMatching})) is given
by
\[\label{eq:solution}
Q^* = U^*+V^*
\]
where $V^*$ is the unique minimizer of 
\begin{equation}\label{eq:VQuad}
\|\P_{\mathcal{X}}\left(W_L^{-1}VW_R^{-1})\right\|_{\Htwo}^2+
2\langle  \P_{\mathcal{X}}\left(W_L^{-1}VW_R^{-1}\right),T
\rangle
\end{equation}
and 
\begin{equation}\label{eq:UFormOpt}
U^* = Q_N
-W_L\P_{\frac{1}{z^{N+1}}\Htwo}\left(W_L^{-1}V^*W_R^{-1}\right)W_R.
\end{equation}
Here $Q_N$ is the optimal centralized delayed controller from Theorem
\ref{thm:centralizedDelayed}. 
}
\end{theorem}

\gap

The theorem can be proved by combining the following two lemmas:

\gap

\begin{lemma}\label{lem:U}
{\it 
For any $V\in\mathcal{Y}$, the optimal solution to
\begin{equation}\label{eq:auxModelMatching}
\begin{array}{c}
\min_{U}\|P_{11}+P_{12}VP_{21}+P_{12}UP_{21}\|_{\Htwo} \\
\textrm{s.t. } U\in \frac{1}{z^{N+1}}\Htwo
\end{array}
\end{equation}
is given by 
\begin{equation}\label{eq:UForm}
U(V) = Q_N
-W_L\P_{\frac{1}{z^{N+1}}\Htwo}\left(W_L^{-1}VW_R^{-1}\right)W_R.
\end{equation}
with optimal cost given by 
\begin{eqnarray}
\nonumber
\lefteqn{
\|P_{11}+P_{12}VP_{21}+P_{12}U(V)P_{21}\|_{\Htwo}^2 
}
\\&=&
\label{eq:decompCost}
\|P_{11}+P_{12}Q_NP_{21}\|_{\Htwo}^2+
\\ &&
\nonumber
\|\P_{\mathcal{X}}\left(W_L^{-1}VW_R^{-1})\right\|_{\Htwo}^2+
2\langle  \P_{\mathcal{X}}\left(W_L^{-1}VW_R^{-1}\right),T
\rangle
\end{eqnarray}
}
\end{lemma}

\gap

\begin{lemma}\label{lem:V}
{\it 
The expression in Equation (\ref{eq:VQuad}) 
has a unique minimum $V^*$ which can be efficiently computed by
quadratic programming. 
}
\end{lemma}

\gap

\begin{remark}
Note that Equation (\ref{eq:UForm}) implies that the optimal $U$ is always the sum of the optimal delayed controller, $Q_N$, and a correction term that depends linearly on $V$. Furthermore, Equation (\ref{eq:decompCost}) shows that optimal decentralized cost is the cost of the delayed controller minus benefits gained from choosing $V$. In particular if $V=0$, then the delayed cost is recovered. 
\end{remark}

\gap

To see how the lemmas prove Theorem \ref{thm:main}, assume that $U^*$
and $V^*$ are optimal. By optimality, $U^*$ must solve Problem
(\ref{eq:auxModelMatching}) with $V=V^*$. Thus Equation
(\ref{eq:UFormOpt}) holds. Furthermore, optimality of
$V^*$ implies that it must minimize the right side of Equation
(\ref{eq:decompCost}), which is equivalent to minimizing Equation
(\ref{eq:VQuad}). 

To complete the proof of Theorem \ref{thm:main}, the lemmas will now
be proved.

\gap

\begin{proof}[Lemma \ref{lem:U}]
First Equation (\ref{eq:UForm}) will be derived, and then the form
will be used to derive Equation (\ref{eq:decompCost}).
If $U$ solves Problem (\ref{eq:auxModelMatching}), then, as in the
proof of the centralized delayed case (Theorem
\ref{thm:centralizedDelayed}), a necessary condition for optimality is
given by 
\[\label{eq:optimalityAux}
\begin{matrix}
P_{12}^{\sim}P_{11}P_{21}^{\sim}+P_{12}^{\sim}P_{12}VP_{21}P_{21}^{\sim}
\\
+
P_{12}^{\sim}P_{12}UP_{21}P_{21}^{\sim} 
\end{matrix}
\in \frac{1}{z^{N+1}}\Htwo^{\perp}. 
\]
Plugging in the spectral factorizations shows that 
\[
\begin{matrix}
P_{12}^{\sim}P_{11}P_{21}^{\sim}+W_L^{-\sim}W_L^{-1}VW_R^{-1}W_R^{-\sim}
\\
+W_L^{-\sim}W_L^{-1}UW_R^{-1}W_R^{-\sim}
\end{matrix}
\in\frac{1}{z^{N+1}}\Htwo^{\perp}.
\]
By anticausality, multiplying on the left and right by $W_L^{\sim}$ and $W_R^{\sim}$,
respectively, gives,
\[
\begin{matrix}
W_L^{\sim}P_{12}^{\sim}P_{11}P_{21}^{\sim}W_R^{\sim}
\\
+
W_L^{-1}VW_R^{-1}+W_L^{-1}UW_R^{-1} 
\end{matrix}
\in \frac{1}{z^{N+1}}\Htwo^{\perp}.
\]
As in the centralized delayed case,
$W_L^{-1}UW_R^{-1}\in\frac{1}{z^{N+1}}\Htwo$ and the left side can be
set to zero by projection:
\begin{eqnarray*}
\lefteqn{
\begin{array}{l}
\P_{\frac{1}{z^{N+1}}\Htwo}\left(
W_L^{\sim}P_{12}^{\sim}P_{11}P_{21}^{\sim}W_R^{\sim}
\right)
\\
+
\P_{\frac{1}{z^{N+1}}\Htwo}\left(
W_L^{-1}VW_R^{-1}+
W_L^{-1}UW_R^{-1}\right) 
\end{array}
}
 \\
&=&  \P_{\frac{1}{z^{N+1}}\Htwo}(T)
\\ &&
+\P_{\frac{1}{z^{N+1}}\Htwo}\left(
W_L^{-1}VW_R^{-1}
\right)
+W_L^{-1}UW_R^{-1} 
\\ &=& 0.
\end{eqnarray*}
Rearranging and multiplying on the left and right by $W_L$ and $W_R$, gives the form of $U$:
\begin{eqnarray*}
U &=& -W_L\P_{\frac{1}{z^{N+1}}\Htwo}(T)W_R
\\ &&
-W_L\P_{\frac{1}{z^N}\Htwo}\left(W_L^{-1}VW_R^{-1}\right)W_R \\
&=& Q_N -W_L\P_{\frac{1}{z^{N+1}}\Htwo}\left(W_L^{-1}VW_R^{-1}\right)W_R,
\end{eqnarray*}
where $Q_N$ is the solution from Theorem
\ref{thm:centralizedDelayed}. Thus Equation (\ref{eq:UForm}) has been proved.

Now Equation (\ref{eq:decompCost}) must be proved. 
The full controller, $Q$, is given by 
\begin{equation}\label{eq:QfromV}
Q = Q_N
-W_L\P_{\frac{1}{z^{N+1}}\Htwo}\left(W_L^{-1}VW_R^{-1}\right)W_R
+V.
\end{equation}
The second and third terms can be expressed as
\begin{eqnarray}\label{eq:preG}
\lefteqn{
-W_L\P_{\frac{1}{z^{N+1}}\Htwo}\left(W_L^{-1}VW_R^{-1}\right)W_R
+V
}
\\
\nonumber
&=&
W_L\left(
W_L^{-1}VW_R^{-1}-\P_{\frac{1}{z^{N+1}}\Htwo}\left(W_L^{-1}VW_R^{-1}\right)
\right)
W_R 
\\
\nonumber
&=& 
W_L\left(\left(
\P_{\frac{1}{z}\Htwo}-\P_{\frac{1}{z^{N+1}}\Htwo}
\right)
(W_L^{-1}VW_R^{-1})
\right)W_R
\\
\nonumber
&=& 
W_L\P_{\mathcal{X}}\left(
W_L^{-1}VW_R^{-1}
\right)W_R 
\end{eqnarray}
Note that the third equality follows since $\P_{\mathcal{X}} =
\P_{\frac{1}{z}\Htwo} - \P_{\frac{1}{z^{N+1}}\Htwo}$. Defining $G$ by 
\begin{equation}\label{eq:G}
G=\P_{\mathcal{X}}\left(
W_L^{-1}VW_R^{-1}
\right),
\end{equation}
the controller $Q$ can now be
written as
\begin{equation}\label{eq:QfromG}
Q=Q_N+W_RGW_L.
\end{equation}

Plugging Equation (\ref{eq:QfromG}) into 
$\|P_{11}+P_{12}QP_{21}\|_{\Htwo}^2$ gives a quadratic function of
$G$:
\begin{eqnarray}
\nonumber
\lefteqn{
\|P_{11}+P_{12}QP_{21}\|_{\Htwo}^2 
}
\\
\nonumber
&=& 
\|
P_{11}+P_{12}(Q_N+W_LGW_R)P_{21}\|_{\Htwo}^2
\\
\nonumber
&=&
\|P_{11}+P_{12}Q_NP_{21}\|_{\Htwo}^2
+\|P_{12}W_LGW_RP_{21}\|_{\Htwo}^2
\\
\label{eq:costExpansion}
&&
+
2
\langle
P_{11}+P_{12}Q_NP_{21},P_{12}W_LGW_RP_{21}
\rangle,
\end{eqnarray}
where the third term is real because $Q_N$ and $G$ must have real
coefficients.   

The second term of Equation
(\ref{eq:costExpansion}) can be simplified as
\begin{eqnarray}
\nonumber
\lefteqn{
\langle
P_{12}W_LGW_RP_{21},P_{12}W_LGW_RP_{21}
\rangle
}
\\
\nonumber
&=&
\langle
W_L^{\sim}P_{12}^{\sim}P_{12}W_LGW_RP_{21}P_{21}^{\sim}W_R^{\sim},G
\rangle
\\
\nonumber
&=&
\langle
W_L^{\sim}W_L^{-\sim}W_L^{-1}W_LGW_RW_R^{-1}W_R^{-\sim}W_R^{\sim},G
\rangle
\\
&=&
\label{eq:costExpTerm3}
\langle
G,G
\rangle
\end{eqnarray}

Similarly, the third term of Equation (\ref{eq:costExpansion}) can be simplified
as  
\begin{eqnarray}
\nonumber
\lefteqn{
\langle
P_{11}+P_{12}Q_dP_{21},P_{12}W_LGW_RP_{21}
\rangle 
}
\\
\nonumber
&=& 
\langle
W_L^{\sim}P_{12}^{\sim}P_{11}P_{21}^{\sim}W_R^{\sim},G
\rangle
\\
&& \nonumber
+
\langle
W_L^{\sim}P_{12}^{\sim}P_{12}Q_NP_{21}P_{21}^{\sim}W_R^{\sim},G
\rangle 
\\
&=& 
\nonumber
\langle
T,G
\rangle
+
\langle
W_L^{\sim}W_L^{-\sim}W_L^{-1}Q_NW_R^{-1}W_R^{-\sim}W_R^{\sim},G
\rangle 
\\
\nonumber
&=&
\langle
T,G
\rangle
+
\langle
W_L^{-1}Q_NW_R^{-1},G
\rangle
\\
\label{eq:costExpTerm2}
&=&
\langle
T,G
\rangle.
\end{eqnarray}
The fourth equality follows because $G\in \mathcal{X}$ and
$W_L^{-1}Q_NW_R^{-1}\in \frac{1}{z^{N+1}}\Htwo$, which are orthogonal
spaces.  
Combining Equation (\ref{eq:costExpTerm3}) and (\ref{eq:costExpTerm2})
with Equation (\ref{eq:costExpansion})
proves that the cost can be decomposed as 
\begin{eqnarray*}
\lefteqn{
\|P_{11}+P_{12}QP_{21}\|_{\Htwo}^2 
}\\
&=&
\|P_{11}+P_{12}Q_NP_{21}\|_{\Htwo}^2
+\|G\|_{\Htwo}^2+2\langle G,T\rangle.
\end{eqnarray*}
Substituting the definition of $G$ (Equation (\ref{eq:G})), proves
Equation (\ref{eq:decompCost}) and the proof of the lemma is complete. 
\end{proof}

\gap

\begin{proof}[Lemma \ref{lem:V}] Recalling Equation (\ref{eq:G}), $G$ can be expanded as an FIR transfer matrix
\[
G = \sum_{i=1}^N\frac{1}{z^i}G_i
\]
Now the coefficients of $G$ will be computed in terms of $V$, $W_L^{-1}$,
and $W_R^{-1}$.
For notional simplicity, let $H=W_L^{-1}$ and $J=W_R^{-1}$. The
matrices $H$ and $J$ can be expanded as 
\[
\begin{array}{rcccl}
H&=& \sum_{i=0}^{\infty} \frac{1}{z^i}H_i &=& 
\Omega^{1/2}\left(I-\frac{1}{z}\sum_{i=0}^{\infty}\frac{1}{z^i}KA^iB_2
\right) 
\vspace{5pt}
\\
J&=& \sum_{i=0}^{\infty}\frac{1}{z^i}J_i &=& 
\left(
I - \frac{1}{z}\sum_{i=0}^{\infty}\frac{1}{z^i}C_2A^iL
\right) \Psi^{1/2}.
\end{array}
\]
Since $V\in \mathcal{Y}\subset\mathcal{X}$ it can be expanded as
$V=\sum_{i=1}^N\frac{1}{z^i}V_i$.  
It follows that $G_i$ can be written as a linear function of $V$: 
\begin{equation}
\label{eq:GCoef}
G_i = \sum_{
\begin{matrix}
\scriptstyle
j,l \ge 0,\:
k \ge 1 \\
\scriptstyle
j+k+l = i
\end{matrix}
}H_jV_kJ_l.
\end{equation}

Similar to $H$ and $J$, $T$ can be expanded as 
\[
T=\sum_{i=1}^{\infty}\frac{1}{z^i}T_i
=\frac{1}{z}\sum_{i=0}^{\infty}\frac{1}{z^i}\Omega^{1/2}KA^iL\Psi^{1/2}.
\]

The expansions of $G$ and $T$ can now be used to express Equation
(\ref{eq:VQuad}) in a form suitable for numerical evaluation:
\begin{eqnarray}
\nonumber
\lefteqn{
\hspace{-10pt}
\|\P_{\mathcal{X}}\left(
W_L^{-1}VW_R^{-1}
\right)
\|_{\Htwo}^2+2\langle 
\P_{\mathcal{X}}\left(
W_L^{-1}VW_R^{-1}
\right)
,
T\rangle 
}
\\ 
\nonumber
&=& \|G\|_{\Htwo}^2+2\langle G,T\rangle
\\
\label{eq:quadExpansion}
&=&
\sum_{i=1}^N\Tr\left(
G_iG_i^T
\right)+
2\sum_{i=1}^N\Tr\left(
G_iT_i^T
\right).
\end{eqnarray}

Note that Equations (\ref{eq:GCoef}) and (\ref{eq:quadExpansion}) can
be used to define a convex quadratic program in $V$. 
If the quadratic form $\sum_{i=1}^N\Tr\left(
G_iG_i^T
\right)$ is positive definite in $V$, then the right side of Equation
(\ref{eq:quadExpansion}) must have a unique minimum which is
efficiently computable. 

The proof can thus be completed by showing that $\|G\|_{\Htwo}^2=0$ implies
that $V=0$. Assume that $\|G\|_{\Htwo}^2=0$. By the positive
definiteness of norms, it must be that $G=0$. Equations
(\ref{eq:preG}) and (\ref{eq:G}) imply that 
\[
W_LGW_R = V-W_L\P_{\frac{1}{z^{N+1}}\Htwo}\left(W_L^{-1}VW_R^{-1}\right)W_R,
\]
and thus, by projection,
\[
V = \P_{\mathcal{X}}\left(
W_LGW_R
\right).
\]
Therefore, $G=0$ implies that $V=0$ and the proof is complete. 
\end{proof}

%% file: examples.tex
\section{Numerical Examples}
\label{sec:examples}

The results in this paper demonstrate that decentralized model
matching with communication delays can be efficiently solved by in terms of the original state matrices. In particular, aside from centralized Riccati 
equations, the only numerical computation required is a quadratic
program specified by Equations (\ref{eq:GCoef}) and
(\ref{eq:quadExpansion}). This section demonstrates the method with a
few examples.

\subsection{The Chain Problem}

\begin{figure}
\centering
\includegraphics[width=.4\columnwidth]{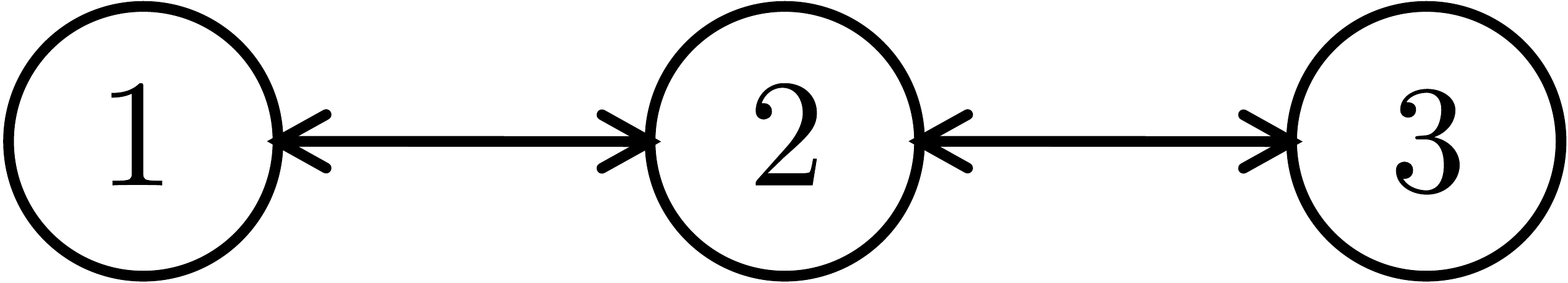}
\caption{\label{fig:chainGraph}
The graph depicts the the communication structure of the three-player
chain problem. Players $1$ and $3$ pass information to player $2$
after a single step delay, while player $2$ passes information to players
$1$ and $3$ after a single step of delay.
}
\end{figure}

The three-player
chain structure, \cite{lamperskistructure2011}, is a delayed
information sharing pattern specified by the graph in 
Figure \ref{fig:chainGraph}. In the frequency domain, the information
structure is represented by the constraint $\mathcal{K}\in \mathcal{S}_{\mathrm{Ch}}=
\mathcal{Y}_{\mathrm{Ch}}\oplus \frac{1}{z^3}\mathcal{R}_p$, where
$\mathcal{Y}_{\mathrm{Ch}}$ is given in Equation
(\ref{eq:chain}). Consider the plant specified by 
\begin{eqnarray*}
A&=& 
\begin{bmatrix}
0.5 & 0.2 & 0 \\
0.2 & 0.5 & 0.2 \\
0 & 0.2 & 0.5
\end{bmatrix},
\\
B&=& 
\left[
\begin{array}{cc:c}
I_{3\times 3} & 0_{3\times 3} & I_{3\times 3}
\end{array}
\right], 
\\
C &=& 
\left[
\begin{array}{c}
I_{3\times 3} \\
0_{3\times 3} \\
\hdashline
I_{3\times 3}
\end{array}
\right],
\\
D &=& 
\left[
\begin{array}{cc:c}
0_{3\times 3} & 0_{3\times 3} & 0_{3\times 3} \\
0_{3\times 3} & 0_{3\times 3} & I_{3\times 3}\\
\hdashline
0_{3\times 3} & I_{3\times 3} & 0_{3\times 3}
\end{array} 
\right].
\end{eqnarray*}

For comparison purposes, the
optimal $\mathcal{H}_2$ norm was computed using model matching from
this paper,  the LMI method of 
\cite{rantzerseparation2006,gattamigeneralized2006}, and the
vectorization method of \cite{rotkowitzcharacterization2006}. In all three cases the norm
was found to be $2.1082$. In contrast, the centralized
controller, $Q_0$, gives a norm of  $2.0853$, while the delayed controller, $Q_2$,
gives a norm of $2.1780$. This is to be expected, since the
controller obeying the three-player chain structure is more
constrained than $Q_0$, but less constrained than $Q_2$:
$\frac{1}{z^3}\Htwo\subset
\left(
\mathcal{S}_{\mathrm{Ch}}\cap\frac{1}{z}\Htwo
\right)
\subset \frac{1}{z}\Htwo$.

\subsection{Increasing Delays}

\begin{figure}
\centering
\includegraphics[width=.9\columnwidth]{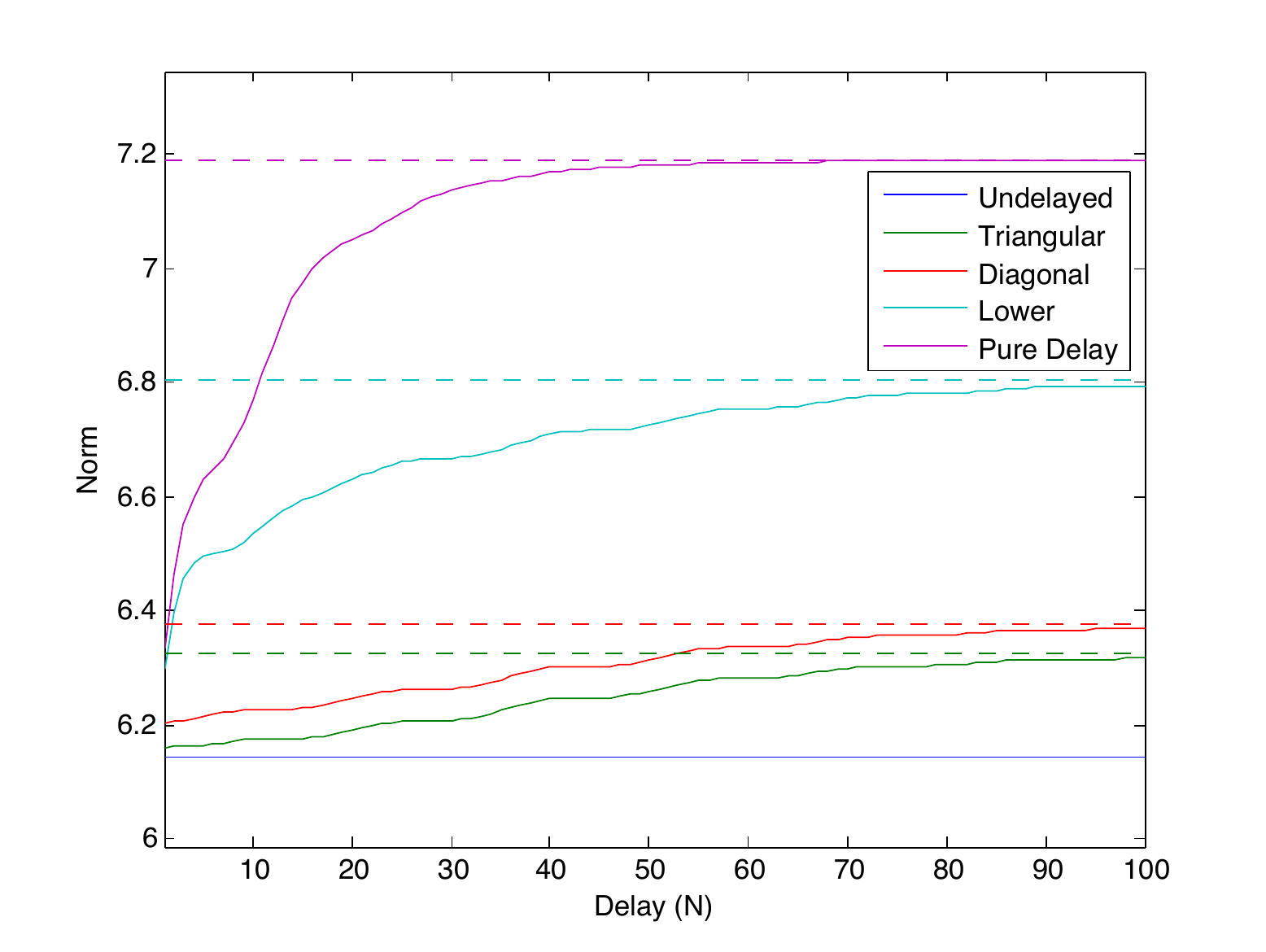}
\caption{\label{fig:exp} This plot shows the closed-loop norm for
 $Q_{\mathrm{Tri}}^N$, $Q_{\mathrm{Di}}^N$, $Q_{\mathrm{Low}}^N$, and
  $Q_N$ (the pure delay case). For a given $N$, the controllers with
  fewer sparsity constraints give rise to lower norms.  
As $N$ increases, all of the norms
  increase monotonically since the controllers have access to less
  information. The dotted lines correspond to the optimal norms for
  sparsity structures given in 
  Equation (\ref{eq:sparseController}). For pure delay, $Q_N\to 0$ as
  $N\to \infty$, and thus the norm approaches the open-loop value. 
}
\end{figure}

Consider the plant with matrices given by
\[
A =
\begin{bmatrix}
1 & 0.2 & 0 & 0\\
-0.2 & 0.8 & 0 & 0.2\\
0 & 0 & 1 & 0.2 \\
0& -0.2 & -0.2 & 0.8
\end{bmatrix}, 
\]
\[
B =  
\left[
\begin{array}{cccc:cc}
0 & 0 & 0 & 0  & 0 & 0 \\
0.2& -0.2 & 0 & 0 & 0.2 & 0\\
0 & 0 & 0 & 0  & 0 & 0 \\
0.2 & 0.2 & 0 & 0 & 0 & 0.2
\end{array}
\right],
\]
\[
C = 
\left[
\begin{array}{cccc}
10 & 0 & -10 & 0  \\
0& 0 & 0 & 0 \\
0 & 0 & 0 & 0 \\
\hdashline
1 & 0 & 0 & 0 \\
0 & 0 & 1 & 0
\end{array}
\right],
\]
\[
D =
\left[
\begin{array}{cccc:cc}
0 & 0 & 0 & 0 & 0 & 0 \\
0 & 0 & 0 & 0 & 1 & 0 \\
0 & 0 & 0 & 0 & 0 & 1 \\
\hdashline
0 & 0 & 1 & 0 & 0 & 0 \\
0 & 0 & 0 & 1 & 0 & 0
\end{array}
\right].
\]

For $N\ge 1$, let $Q_{\mathrm{Tri}}^N$, $Q_{\mathrm{Di}}^N$, and $Q_{\mathrm{Low}}^N$ solve the decentralized model matching problem, Equation (\ref{eq:modelMatching}), with the form 
\begin{eqnarray*}
Q_{\mathrm{Tri}}^N &=& U_{\mathrm{Tri}}^N+V_{\mathrm{Tri}}^N, \\
Q_{\mathrm{Di}}^N &=& U_{\mathrm{Di}}^N+V_{\mathrm{Di}}^N, \\
Q_{\mathrm{Low}}^N &=& U_{\mathrm{Low}}^N+V_{\mathrm{Low}}^N.
\end{eqnarray*}
Here $U_{\mathrm{Tri}}^N$, $U_{\mathrm{Di}}^N$, $U_{\mathrm{Low}}^N\in \frac{1}{z^{N+1}}\Htwo$ and $V_{\mathrm{Tri}}^N$, $V_{\mathrm{Di}}^N$, $V_{\mathrm{Low}}^N$ are FIR transfer matrices with sparsity structure given by
\begin{eqnarray*}
V_{\mathrm{Tri}}^N &=& \sum_{i=1}^N \frac{1}{z^i} 
\begin{bmatrix} 
* & 0 \\
* & *
\end{bmatrix}, 
\\
V_{\mathrm{Di}}^N &=& \sum_{i=1}^N \frac{1}{z^i}
\begin{bmatrix}
* & 0 \\
0 & *
\end{bmatrix},
\\
V_{\mathrm{Low}}^N &=& \sum_{i=1}^N \frac{1}{z^i}
\begin{bmatrix}
0 & 0 \\
0 & *
\end{bmatrix}.
\end{eqnarray*}
The resulting norms are plotted in Figure (\ref{fig:exp}).

As $N\to \infty$, the resulting controllers appear to approach optimal sparse
controllers 
\begin{eqnarray}
\nonumber
Q_{\mathrm{Tri}}^{\infty} &\in& \begin{bmatrix}
\frac{1}{z}\Htwo & 0 \\
\frac{1}{z}\Htwo & \frac{1}{z}\Htwo
\end{bmatrix} 
\\
\label{eq:sparseController}
Q_{\mathrm{Di}}^{\infty} &\in & 
\begin{bmatrix}
\frac{1}{z}\Htwo & 0 \\
0 & \frac{1}{z}\Htwo
\end{bmatrix}
\\
\nonumber
Q_{\mathrm{Low}}^{\infty}&\in & 
\begin{bmatrix}
0 & 0 \\
0 & \frac{1}{z}\Htwo
\end{bmatrix},
\end{eqnarray}
which can be computed by the vectorization technique from
\cite{rotkowitzcharacterization2006}. Evidence for the convergence is
shown by the fact that the norms limit to the values computed for the
sparse controllers (Figure \ref{fig:exp}).

%% file: conclusion.tex
\section{Conclusion}
\label{sec:conclusion}

This paper derives a novel solution for a class of output feedback
$\Htwo$ model matching problems with communication delays.   
To find the optimal solution, the controller is decomposed into
orthogonal components, both of which are easily computable. In
particular, centralized delayed controllers that optimally correct for
the FIR component are computed by spectral factorization. Then, the
problem is then reduced to 
optimization over the FIR component. 

The results of this paper indicate that the optimal control can be
computed in terms of the centralized Riccati equations for the system. Existing
time-domain methods, such as
\cite{nayyardecentralized2012,rantzerseparation2006,gattamigeneralized2006},
work with state variables that have been augmented to include memory
vectors required by the various controllers. The optimal controllers are then
constructed based on centralized solutions to the augmented-state
problems. It would be interesting to see if these alternative
constructions can be mapped onto one another. In particular, the
augmented-state solutions could lend insight into the computation of
the FIR terms, while the method of this paper might be used to
construct solutions to the augmented-state problems in terms of
optimal controllers for the original centralized system.
